**Towards a Privacy Research Roadmap for the Computing Community**

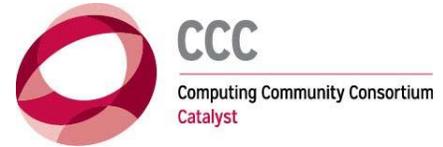

**I. Introduction**

Great advances in computing and communication technology are bringing many benefits to society, with transformative changes and financial opportunities being created in health care, transportation, education, law enforcement, national security, commerce, and social interactions.  Many of these benefits, however, involve the use of sensitive personal data, and thereby raise concerns about privacy.  Failure to address these concerns can lead to a loss of trust in the private and public institutions that handle personal data, and can stifle the independent thought and expression that is needed for our democracy to flourish.

This report, sponsored by the Computing Community Consortium (CCC), suggests a roadmap for privacy research over the next decade, aimed at enabling society to appropriately control threats to privacy while enjoying the benefits of information technology and data science.   We hope that it will be useful to the agencies of the Federal Networking and Information Technology Research and Development (NITRD) Program as they develop a joint National Privacy Research Strategy over the coming months.  The report synthesizes input drawn from the privacy and computing communities submitted to both the CCC and NITRD, as well as past reports on the topic.

Privacy is a broad topic, encompassing a variety of issues in many different contexts.  Our focus is on concerns raised by the collection, sharing, analysis, and use of personal data in information systems.   Even with this bounded scope, the privacy concerns in consideration are manifold, including (but not limited to) unwanted disclosure of personal information, lack of transparency and control around how one's information is used, and discrimination based on personal information.

The research agenda we describe seeks to lead us to a state where:

- We have a rigorous science of privacy that applies across different application domains;
- We understand the needs, expectations, and incentives of the humans who use information systems, and can design systems that are sensitive to them;
- Privacy technology research and privacy policy objectives are informed by and aligned with each other; and
- We can engineer systems that enable us to enjoy both privacy and the benefits of data use to the maximum extent possible, showing that the tradeoff between the two can be much less stark than our current approaches offer:

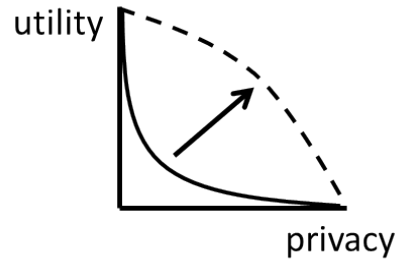

To reach this state, we believe that the research strategy needs to:

- *Emphasize understanding, defining, and measuring the privacy of information systems:* A major challenge for privacy research is that there is no single, agreed-upon definition of "privacy"; indeed, the term refers to a variety of distinct concerns. However, for each particular privacy concern, it is important to understand and precisely define the objective, evaluate it with scientific rigor, and match it to the requirements of particular application domains. This is a prerequisite for transparency and privacy-by-design that should be a priority for privacy research more broadly.

- *Recognize and support the many stages and dimensions of privacy research:* The pipeline in privacy research starts from foundational work that aims to understand phenomena and the range of technological possibilities and limitations, continues on to applied research that is directed at specific privacy objectives, and finally to translational work that seeks practical impact on particular application domains. Support is needed for research at all of these stages, as well as for ensuring a steady flow of ideas from each to the next.

- *Enable interdisciplinary research strategies:* Developing a science of privacy and effective privacy solutions requires a combined understanding of computing technology, information, human behavior, and governance mechanisms. Thus, it is important to develop and strengthen structures that encourage interdisciplinary research on privacy. At the same time, it should be recognized that multi-disciplinary approaches are also appropriate and that some research directions need to mature within a single field before crossing disciplinary lines.

- *Foster a technology-policy dialogue:* We need a bi-directional conversation that enables regulators, lawmakers, standards creators, researchers, and system builders to understand what is and is not possible to achieve with technology, and be informed about policy objectives related to privacy properties of systems, procedures, and processes.

Process Leading to this Report

In Spring 2014, the White House and the President's Council of Advisors on Science and Technology (PCAST) conducted 90-day reviews of big data and privacy.[1]  Both reports recommend an increased federal investment in privacy research.  Consequently the NITRD agencies were tasked with developing a National Privacy Research Strategy (NPRS), which should "establish objectives and prioritization guidance for federally-funded privacy research, provide a framework for coordinating R&D in privacy-enhancing technologies, and encourage multi-disciplinary research that recognizes the responsibilities of the Government, the needs of society, and enhances opportunities for innovation in the digital realm."  To gather community input for the NPRS, NITRD issued a Request for Information in September 2014 and held a workshop in February 2015.   Following the workshop, the Computing Community Consortium (CCC) formed a committee of five researchers to gather additional community input and synthesize it into a coherent research agenda.  This report is the result of that effort.  It draws upon the Big Data Privacy reports, the responses to the September 2014 RFI, the February 2015 workshop, and additional community input gathered by the committee.  While the White House and PCAST Big Data Privacy reports explain the privacy implications of big data and information systems, and near-term policy and technology solutions, the current report has a somewhat wider scope (all uses of personal data in information systems, not just "big data"), and focuses on objectives for a longer timescale (e.g. a decade) and the research activities needed to reach those objectives.

Outline of the Remainder of the Report:

- Section II lists a number of national priority areas and application domains where information systems and the use of personal data raises opportunities for great benefit as well as threats to privacy.
- Section III outlines several broad privacy objectives that cut across these application domains.
- Section IV presents and organizes many specific research areas that, together, can help us reach the privacy objectives stated in Section III.
- Section V offers suggestions for encouraging the much-needed interdisciplinary research on privacy.
- Section VI discusses how to enable the outcomes of such research to be translated into practice so as to serve the national priorities and application domains of Section II.

---

[1] "Big Data: Seizing Opportunities, Preserving Values," May 2014, http://www.whitehouse.gov/sites/default/files/docs/big_data_privacy_report_may_1_2014.pdf.
"Big Data and Privacy: A Technological Perspective," May 2014, http://www.whitehouse.gov/sites/default/files/microsites/ostp/PCAST/pcast_big_data_and_privacy_-_may_2014.pdf.

## II. Domains

This section describes a variety of national priorities and application domains for which privacy research is important, detailing the benefits of the use of personal data and the risks involved in each. As discussed in the introduction, the privacy research agenda should seek to develop a science of privacy that cuts across application domains. There is also a need for supporting more applied, domain-specific research that translates the broad scientific knowledge into practical impact within the specific domains.

A. Health Care
Information systems have the potential to vastly improve health care services; increased sharing and use of health and medical data will produce a variety of benefits for this sector, including more accurate diagnosis, more personalized and coordinated care, more rapid development of new treatments, faster response to distress via novel health-tracking devices, more effective treatment based on genetic makeup, and lower costs. Privacy concerns include disclosure of sensitive health data through sharing or data breaches, discrimination in employment or insurance based on medical conditions or even genetic predisposition, and continual monitoring of patients outside the healthcare context through novel health-tracking devices.

B. Transportation
Benefits of information technology on transportation can be in reducing congestion, preventing accidents, reducing deaths and injuries, increasing fuel efficiency, and saving human effort spent on driving. Privacy concerns come from the tracking of individual movements through navigation systems, roadway sensors, traffic cameras, in-car data collection, and communications between cars.

C. Law Enforcement and National Security
Law enforcement and intelligence agencies collect and analyze many different types of data (criminal records and non-criminal supplemental information) to create a "virtual picture" of individuals to help with solving crimes, preventing attacks, and tracking terrorists. The concern is that these organizations collect information en-masse on the general population, raising the possibility of unauthorized use and the chilling effects of surveillance.

D. Education
Information technology and data can improve education by enabling adaptive and personalized content, measurement of teacher performance, more effective and efficient education policy, and tools to assist families in educational choices. Online courses provide vast increases in the availability of education for all learners. Privacy concerns arise from the sensitivity of data about students' engagement and performance, including exposure of student home study habits as recorded by online interfaces.

E. Modern Internet Services
Modern Internet services, such as search engines, social networks, online video services, and online retailers, have access to a rich array of data that can be used for useful purposes, including advanced personalization of content, novel forms of social interaction, and connecting people to other people, businesses and government. There are also opportunities to use such data for research and other socially beneficial purposes (e.g. prevention of disease outbreaks, aid in natural disasters). Concerns arise from use, abuse, and sharing of the data for purposes other than the ones for which it was provided, and unfairness or discrimination that can arise from personalized services.

F. Modern App Ecosystems

Devices such as smartphones, web browsers, activity bracelets and their apps provide great utility, entertainment, and functionality to users. The generative ecosystem that has evolved for easily creating and distributing apps has led to great innovation, but makes it a challenge to ensure that apps respect users' privacy and security. In particular, it is difficult for users to control the apps' access to the vast amounts of sensitive data stored on the devices and to the various sensors that are built into the devices.

G. Internet of Things and Smart Infrastructures

The internet of things and smart infrastructure – smart buildings, smart homes, smart cities – enable improvement of living conditions, productivity, and quality of life. For example, a smart home can detect the occupants present, learn their schedules and requirements, and then combine that information with real-time utility prices and smart meters to provide efficient and automatic control of appliances. However, the same information can be used to track when individuals are home, which TV programs they watch and websites they visit, their sleep schedules, and other behavior. The risk is the exploitation of such data for other purposes such as lawsuits, insurance decisions, unwanted advertising, or crime.

H. Financial Sector

Data from financial institutions can assist regulators in assessing compliance, and allow for analysis of trends and forewarn about such dangers as the 2009 financial crisis. However, financial data is sensitive not only at the level of individual customers, but also at the level of institutions, since it reveals proprietary information about strategies and holdings.

I. Open Government Data

Governments at all levels are releasing large amounts of data in order to increase trust and transparency and to enable innovative applications. However, these data releases often relate to sensitive information about citizens, creating a tension between governments' obligations to share data and to respect privacy.

J. Research Data

As the advancement of knowledge in many fields becomes increasingly data-driven, it is important that the data supporting research results be made available for validating and extending the findings, and indeed many funding agencies (such as NSF and NIH) have adopted policies that mandate sharing of research data.  However, for data on human subjects, researchers currently lack adequate tools to protect the privacy of their subjects when sharing their data; the traditional approach of deidentification by removing identifiers is now known to provide very weak protections.

## III. Privacy Objectives and Desired Capabilities

Privacy research has important contributions to make to society in the quest to design, operate and regulate information systems that respect a range of privacy values.  First, we depend on progress in computer science, social science, and law to enhance our ability to protect society's established privacy values.  Second, researchers in a variety of fields have much to contribute to society's ongoing understanding of the value, meaning and practical application of privacy in our information-intensive world.  Privacy is neither a unitary nor absolute quality.  Some uses of personal data may both offer important economic, scientific, or social value, but at the same time pose privacy risks.  So engagement with scholars in a variety of fields is especially important to help society make decisions and find the appropriate alignment of privacy rights along with other social values.

Computing researchers, both working within their disciplines and in increasingly important interdisciplinary research projects, can help achieve the key privacy policy objectives identified in the White House Consumer Privacy Bill of Rights and the Federal Trade Commission report "Protecting Privacy in an Era of Rapid Change," namely: data transparency, accountable information use, respect for context, individual control over personal data, privacy-by-design, access and correction to personal data, fairness, and security.[2]

To meet these objectives, research on privacy should aim to yield the following capabilities in the future: measurement of privacy, social science of privacy, security for privacy, engineering of privacy, and policy for privacy.  These capabilities, and the privacy objectives they support, are important for all of the application domains mentioned in Section II.   We now elaborate on each of these desired capabilities:.

- Measurement of Privacy: *we should be able to precisely define what various privacy objectives mean for different types of information systems, and be able to measure the extent to which the systems meet those objectives.*

    Measuring the privacy behavior of information systems supports many privacy objectives, including individual control, accountability, respect for context, and

---

[2] This is an amalgam of the key principles and goals identified by the White House and the FTC.

transparency. In order to ensure that a system provides "individual control," we must identify the presence of personal data in a system such that it can be controlled. To make collection and use of personal data accountable to relevant rules, we must have some visibility into the behavior of these systems. Whether a system respects the context in which information was originally collected can only be determined by measuring the state of the system at the time of collection and then throughout the use of that data.

Measurement is particularly important to achieve the policy objective of *transparency*. If the collection, flow and use of personal data is not at least partially visible either to a person or a machine, then there is no way to assess or control the privacy consequences of any system. In the past, the policy requirement of transparency, found in all leading privacy legal instruments,[3] has been understood largely as a value directed at individuals - to enable informed choices about individual privacy relationships. Today we understand that individual choice mechanisms may not be effective means of protecting privacy because of imbalances in bargaining power and high transaction costs. Computational measurement capability opens up the possibility of supplementing the already over-taxed cognitive capacity of humans with machine-assisted reasoning and analysis.

Scalable transparency and measurement techniques will be particularly important when systems use personal data in new ways, so that society can develop informed perspectives on the risk and desirability of such uses. Thus, we need tools and techniques that move beyond often-imprecise approximations of what might be happening with personal data in a system, toward more formalized, rigorous and well-founded measures of exposure and risk associated with collection, transfer and use of personal data. Effective privacy measurement techniques will enable us to replace paper-driven transparency mechanisms such as human-readable policies with scalable computational techniques.

In addition, there is a need to enable measurements when there is partial transparency. When companies do not want to disclose their algorithms for proprietary reasons we would still need to be able to determine the privacy exposure. This partial transparency creates a much more challenging setting for solving the measurement problem.

---

[3] Records, Computers, and the Rights of Citizens Report of the Secretary's Advisory Committee on Automated Personal Data Systems, July 1973," May 2015, https://epic.org/privacy/hew1973report/default.html. Privacy Act of 1974, Pub. L. No. 93-579 (5 U.S.C. § 552a). European Union (EU) Data Protection Directive 95/46/EC. OECD Guidelines on the Protection of Privacy and Transborder Flows of Personal Data (1980, revised 2013). http://www.oecd.org/sti/ieconomy/oecdguidelinesontheprotectionofprivacyandtransborderflowsofpersonaldata.htm.

Furthermore, it is vital to investigate the question of whether measuring a system's privacy exposure as a stand-alone entity is sufficient or whether we should take into account the secondary exposures which occur by crossing information with other available content.

- Social Science of Privacy: *we should understand the privacy needs and expectations of the humans who use information systems, the institutional dynamics of the organizations that use personal data, and how larger social and economic forces relate to privacy.*

  Effective privacy protection depends not only on public policy but also on nuanced understanding of human behavior to help system designers tailor their designs to respond to behavioral dynamics. Both policy makers and system designers will depend on insight from behavioral sciences — such as economics, institutional psychology, sociology — to tune both lawmaking and technology design. This is important for achieving all privacy policy objectives, because privacy is ultimately about the effect of the use of personal data on the human users of information systems. But it is particularly important for individual control, access and correction, and respect for context, which refer directly to the interaction of users with information systems, and their expectations when doing so. For example, we need to be able to answer questions such as:

    - How do user interface design patterns influence privacy choices made by users?
    - How do different uses of personal data or system design decisions increase or decrease the risk of chilling effects on individual participation and expression in online environments?
    - How do different privacy laws and social norms regarding personal data change the distribution of the economic value derived from the use of personal data?

- Security for Privacy: *we should understand the relationship between security and privacy, and be able to secure information systems from unauthorized access to personal information.*

  Security is closely related to but not synonymous with privacy. Protecting privacy requires being able to secure personal data from unauthorized access and use. However, while necessary, security alone is hardly sufficient for privacy. Observing the history of privacy intrusions over the last several years is enough to illustrate that simply protecting data from outside intruders is not sufficient to protect privacy, as some privacy harms come from those who are actually authorized to access data in some fashion or another.

Design of secure systems themselves can raise privacy risks and challenging policy trade-offs. Some approaches to detecting and analyzing security threats depend on collecting larger volumes of personal data. Of course, securing systems is important to privacy protection. At the same time, more research is required to learn how to collect threat intelligence and attribution data without compromising privacy interests of innocent users of systems under investigation.

While computer security has been studied for a long time, many breaches of confidential personal data still come from failures of security. The causes of these failures include issues with the usability of security technology and the increasing complexity of the systems that we need to secure, and thus continued research on computer security and the transition of its results to practice remains important for privacy. Furthermore, there is a need to examine whether we can tolerate solutions that present a loss in efficiency but provide better privacy guarantees.

- Engineering of Privacy: *we should be able to design and build information systems that meet privacy objectives while allowing us to enjoy the beneficial uses of personal data.*

    Privacy policy makers and scholars have called for *privacy-by-design* — techniques to build privacy protection into the initial design of systems, rather than trying to retrofit designs or manage privacy purely through human-controlled policies. Large numbers of developers are now designing systems that handle personal data. These developers need well-designed, reliable, and modular software, hardware, and services that have privacy awareness built in. For example, we need tools that allow for data analysis while minimizing exposure or inappropriate use of personal data, that log use of personal data and assess those uses against formally specified rules, that track the provenance and onward flows of personal data, and that offer individual users meaningful transparency and control of how their data is used. Hardware design, especially in embedded systems that are difficult to modify as well as in general-purpose hardware technologies that support protected execution environments, may also have privacy impact and can benefit from advance thinking about personal information handling as well.

    Being able to engineer information systems that enable meeting the many

privacy policy objectives requires significant advances in basic research (including on measurement of privacy, the social science of privacy, and security, as discussed above, and the many other areas discussed in Section IV below), as well as substantial investment in the transition to practice (discussed in Section VI below).

- Policy for Privacy: *we should be able to design effective laws, regulations, policies, and best practices regarding the use of personal data in information systems in a way that recognizes the unique capabilities and limitations of information systems.*

    Neither technology nor policy on its own will be effective in meeting privacy objectives.  Just as technological systems need to be engineered with an understanding of the policy objectives, privacy laws, rules, regulations and best practices need to be developed with an understanding of what is technologically possible and impossible.  To guide the development of a broad policy approach to privacy objectives, we need research that can answer questions such as these:

    - How are different legal and regulatory approaches likely to meet privacy objectives given the unique nature of today's information systems and the global reach of many information infrastructures?
    - What can we learn about developing privacy policy from other regulatory areas such as environmental policy, human rights law, financial services regulation, and telecommunications regulation?
    - What can we learn from the way actual institutions are implementing the wide variety of existing privacy policies?
    - What are the limits of technology and what would need to be enforced by laws?

    Answers to these kinds of questions will require an ongoing dialogue between policymakers, technologists, practitioners, and scholars from a variety of disciplines.  Policymakers at regulatory agencies such as the FTC and FCC should find ways to engage technologists and scholars in research that will inform policy decisions, perhaps working with funding agencies such as NSF.

**IV. Research Directions**

In this section, we describe a number of research areas where substantial advances are needed to reach the capabilities discussed in Section III.  Collectively, these research directions involve a wide variety of disciplines, which include computing and mathematical sciences, social sciences, and law and policy.   Meeting privacy policy objectives requires not only advances within the individual disciplines, but a significant amount of interdisciplinary work that integrates insights from multiple areas.  Similarly, we need both basic research to develop a science of privacy that cuts across application domains, as well as applied research that translates this science into practical tools that fit the particulars of a given application.  Sections V and VI of the report provide suggestions on how to foster interdisciplinary work and enable the transition to practice, respectively.

A. Definitions and Frameworks

A major challenge for privacy research is finding clear and convincing specifications for goals, because privacy refers to a variety of different concerns associated with the use of personal data.  Moreover, it is often difficult to draw a clear distinction between the uses of personal data that we wish to enable in a given context ("utility") and that which we wish to limit ("privacy").

For the success of the privacy research agenda, it is extremely important to overcome these difficulties, and provide precise definitions of different privacy properties in a variety of contexts.  In addition to providing the foundation for a science of privacy, good definitions serve as an interface between policy objectives in application domains and particular solutions or technologies.  That is, they allow us to separate the question of whether the definition meets the needs of particular application domains and policy objectives from whether a particular technology satisfies the definition.   Precise definitions allow the privacy properties of technologies to be evaluated in a scientifically rigorous manner, whether empirically or through mathematical proof.

Some successes of definitional work in privacy are the mathematically rigorous definitions of security and privacy properties in cryptography (such as for secure multiparty computation), and the definition of differential privacy.  These definitions correspond to particular notions of utility (computing arbitrary functions in the case of secure multiparty computation vs. computing global statistical properties in the case of differential privacy), particular notions of privacy (hiding everything except the function output vs. hiding individual-level data), and particular trust models (everyone holds their own data vs. a trusted curator holding all data).  These definitions consider an adversarial threat, but it can also make sense to consider game-theoretic definitions that model the incentives of a potential attacker.

While there have been some such successes, there remain many challenges for definitional work in privacy.  Most privacy objectives currently lack sufficiently precise definitions.  Even in the success cases, the interpretation of the definitions in relation to a policy objective can be subtle (e.g. can revealing the output of a function already allow for a privacy compromise?), reflecting a greater need for engagement between researchers with definitional expertise, policymakers, and domain experts.   Another issue that poses difficulties for many privacy definitions is composability: how does a privacy-protective system that satisfies a particular definition interact with a complex environment, with other systems and sources of data?  Indeed, many unexpected compromises of privacy in the past have come from failure to take into account additional sources of data that might be available to an attacker.

Finding definitions that are mathematically precise, computationally realizable, meet policy objectives, and satisfy the needs of applications can involve numerous disciplines, including computer science, statistics, mathematics, law, philosophy, economics, psychology, and the individual application domains. Research on definitions contributes primarily to developing our capabilities for measurement of privacy.

B. Measurement

To solve the problem, we must understand the problem. In the realm of privacy, this requires new research on measuring how information about individuals is collected and shared today and understanding the impact on the victims of privacy violations. This includes basic research on new analytical and measurement technologies, as well as applied, empirical research on measuring information leakage in deployed systems. In general, it is difficult to measure both the benefits and harms of information sharing. Developing tools that can assess and measure privacy risks and potential harms requires new research.

Specific research topics include design and implementation of techniques for detecting and measuring flows of personal information with only black-box access to systems handling this information and techniques for measuring what is revealed about users by systems that learn from users' data (for example, recommender systems and other systems based on aggregates of personal data). This includes the study of "re-identification attacks" against aggregated data, i.e., any information about a dataset that depends on a large number of individuals, in contrast to tabular datasets where one record corresponds to one individual. A more complete, quantitative understanding of the risks of aggregate data is crucial to advance the field of privacy.

Today, information about individuals is scattered across multiple organizations and data holders. We need research on tools that can help individuals first discover all information pertaining to themselves, and, second, assess whether this information is accurate.

Last, but certainly not least, it is essential to understand the real-world impact of privacy violations on the human victims. Research agendas here could involve better engagement with at-risk communities, such as activists, targets of "doxing" and other cyber bullying, domestic violence victims, etc.

Research on measurement of privacy will benefit from contributions and methodologies from disciplines such as sociology (and other social sciences) and biomedical research.

C. Algorithms, Statistics, and Machine Learning

Many of the benefits of data come from the application of algorithms, statistics, machine learning to gain insights about populations (whether of customers, citizens, medical patients, or research subjects) and to then share or use those insights for the benefit of society, individuals, or companies. This raises several different privacy issues and needs for additional research:

- The statistics or model that results from the analysis can disclose inappropriate information about inputs (i.e. training data) used in the analysis.

  This threat may seem counterintuitive, because a statistical model is supposed to be a result of a population-level analysis, but it has been demonstrated that even seemingly aggregate statistics can reveal detailed information about individuals. (See also Section IV B on Measurement, above.) The past decade of theoretical research on differential privacy has showed that in principle, we can often protect against such threats with little loss to utility. But for this body of work to have a more substantial impact on practice, many theoretical and practical challenges need to be overcome, such as (a) supporting real-world data workflows with cleaning and preprocessing, and iterative model fitting, (b) providing utility guarantees in the language of empirical science (p values and confidence intervals), (c) developing algorithms that provide higher utility, perhaps by relaxing the worst-case notions of privacy and utility that are typically used, (d) managing privacy loss over time (over many data releases and many analysts), and (e) development of programming languages and tools that allow for the easy construction of differentially private algorithms without expert involvement.

- Common approaches to data analysis require collecting all the data in a single place, which incentivizes greater collection and centralization of data, and increases the risk of disclosure through a data breach, subpoena, or other means.

  To address the concern, more research (both theoretical and practical) is needed on privacy-preserving distributed and dynamic data analysis that does not require storing all data in one place. Secure multiparty computation (from Section IV D on Crypotgraphy, below) is a powerful tool for this, which is gradually becoming more practical and can be combined with concepts such as differential privacy. Another approach to avoiding storing lots of sensitive data in one place is to design privacy-protective methods of performing one-shot data releases (after which the original data can be deleted). The traditional approach of anonymization by removing identifiers is now understood to provide very weak privacy protections, so instead we should seek rich statistical summaries (such as generating synthetic datasets) that meet formal privacy definitions such as differential privacy. There are some promising theoretical results along these lines, but more research is needed to make them practical.

- Algorithms that take actions with respect to individuals, for example product recommendations or loan decisions, raise issues of fairness, discrimination, and transparency.

  For example, an algorithm could have "learned" (from its data) to target credit cards with higher interest rates to disadvantaged groups, and little justification for its targeting decisions might be found by inspecting its code (which might simply be a vector of numerical parameters learned during training). Compared to the issues mentioned above, there has been less mathematical and computer science work on this problem so far, so finding good definitions and approaches remains an important challenge for privacy research.

Research on these topics requires interdisciplinary collaboration between theoretical computer scientists, statisticians, empirical data scientists from application domains, computer systems and programming languages researchers, economists, other social scientists, and legal scholars.  The research contributes primarily to our capability for engineering privacy.

D. Cryptography

Cryptography provides many tools for enabling privacy and confidentiality of data, starting with basic cryptographic primitives such as encryption and ranging all the way to advanced functionalities such as secure multiparty computation, privacy-preserving data mining, search on encrypted data, oblivious RAMs, functional encryption, mix nets, anonymous credentials, and at the extreme, new tools such as fully homomorphic encryption and program obfuscation.

Some of the advanced technologies provide privacy by distributing the data, the computation, or the keys.  That is, various parts of the computation elements are held by different parties and via this distribution privacy guarantees are obtained.

Issues that need to be examined include:

- The efficiency and usability of privacy-enhancing cryptographic algorithms.  Design of protocols needs to be done with an eye to efficiency.  Looking at the history of computer security, there were times when users and industry did not want to pay performance costs in order to achieve better security.  Over time, security breaks motivated the introduction of better security, provided that performance costs were reasonable.  In the context of privacy, we must similarly focus on creating the most efficient algorithms possible and providing solutions with good usability.  Furthermore, through interdisciplinary research with social science, we need to understand what users and companies will consider as acceptable tradeoffs in efficiency and usability in order to provide privacy.
- In what cases do cryptographic solutions address issues that are genuinely of concern in application domains?  Answering this question requires interdisciplinary research between cryptographers and domain experts to understand and model both the utility and privacy requirements.  For example is a solution that hides the inputs and the computation but exposes the final result sufficient?  What types of functionality and privacy assurances can be offered via cryptographic tools?  The last point would involve research regarding definitions and measurements of privacy to assess what levels of privacy are offered by any given solution.  Furthermore, there is a need to find better ways to communicate what specific solutions offer.
- How do cryptographic techniques allow law enforcement and intelligence agencies to access only data that they need from a large database?  One approach is to use secure multiparty computation, possibly tailored to the specific queries at hand (e.g. set intersection).  The greatest challenge on this front is to provide solutions that manage to

> balance privacy with national security, and that are compatible with legal oversight mechanisms (such as warrants issued by courts).  This requires interdisciplinary research with law and policy, as well as experts in law enforcement and intelligence.
- Defining models of privacy/information leakage.  In order to achieve sufficient efficiency, we may have to resort to solutions that are not "perfect".  That is, we may need to relax typical theoretical models to obtain practical performance at the  expense of some leakage of information.  Research along these lines needs to define appropriate models, design and analyze protocols and measure their leakage, and determine whether such solutions are still appropriate for deployment.

Cryptography alone is not sufficient for developing robust privacy technologies.  It is an essential piece of the puzzle, but not the only one.  In particular, cryptographic techniques fundamentally depend on the security of the computer systems using them.  Thus, research on cryptography and secure systems should work together towards the development of end-to-end privacy solutions.

E.  Systems

Systems research is a core discipline within computer science that focuses on the design and implementation of computer software and hardware.  Personal data is collected, stored, processed, and analyzed by computing systems. Therefore, it is not possible to meaningfully ensure privacy unless all levels of the system stack are not just "privacy-aware," but incorporate privacy protection at the fundamental design level.

System developers should aim to design and implement systems that provide precisely stated, verifiable privacy guarantees. This requires research on tools and methods for checking and certifying that systems use personal information in ways that respect privacy expectations and (where appropriate) satisfy mathematical properties such as differential privacy or other measures of policy conformance.  Techniques that can help achieve this goal include program analysis methods for various kinds of information flow properties and privacy policy languages that are usable by legal experts, yet have precise semantics that system developers can use to restrict and provide accountability for how their code operates on personal information of users.

If privacy requirements and policies describing permitted and prohibited data uses are specified precisely, trusted hardware can help build trustworthy execution and data handling environments to serve as an alternative to emerging cryptographic tools such as fully homomorphic encryption and obfuscation.

Protecting collected data from abuse requires research on systems that can (1) attach policy conditions to data, (2) track disclosure while ensuring compliance with privacy policies, and (3) track data provenance — where and from whom the data was collected and for what purpose (in cloud-based environments in particular, it is essential to be able to reliably identify owners of information).  Systems handling sensitive data should support accountable systems techniques that enable trustworthy auditing of data collection and data use.

Modern communication systems, networks, and protocols often fail to protect privacy of individuals who use them.  Research is needed on new technologies for anonymous, censorship-resistant, and "metadata-hiding" communications, as well as integrating support for privacy into core Internet infrastructure such as DNS.

Researchers pursuing core systems research should be encouraged to take into account privacy policies and regulations; usability and human factors, which will be critically important when these systems are deployed; and how the data is used in key application domains, including but not limited to social sciences, biology and medicine, economics, and law.  Research on systems aspects of privacy contributes primarily to our capability for engineering of privacy.

F: Usability and human behavior

Individuals often lack awareness about how and by whom their data is being collected, the privacy consequences of their actions, and the tools available to protect themselves.  In addition, usability issues limit the effectiveness of many tools that have been introduced to help individuals protect their digital privacy.  Despite a growing stream of research into usable privacy over the past decade, we have seen only modest improvements in the usability of commercially-available privacy tools.  To improve the usability of privacy tools requires first gaining a better understanding of the mental models, workflows, expectations, and privacy needs of a diverse set of users from different cultures, age groups, and backgrounds.  We also need to study the complexity and diversity of people's privacy preferences and concerns, what people are capable of, and how different privacy interfaces and technologies impact their decisions and behaviors.  Further research is needed on how to codify data practices and user privacy preferences, how to convey data practices and privacy risk to users in a meaningful and actionable way, and how to reduce the burden associated with privacy decision making.  We also need to gather empirical data on the privacy harms that people face in the real world to inform the development of mitigations that will address these actual harms.  Finally, interdisciplinary collaboration is needed to develop more usable privacy tools, grounded in knowledge of human behavior, that leverage machine learning techniques to automate privacy decision making.

We elaborate on a few example directions for research in this area:

- *Cognitive and behavioral analysis of privacy decisions.* Further research is required to better understand cognitive and behavioral biases and other factors related to privacy decisions and develop tools and approaches that take into consideration, or even counter, those factors which may bias decision-making. Privacy decisions are difficult for people to make for several reasons, including incomplete and asymmetric information regarding data usage, and limited mental resources to evaluate all possible options and consequences of their actions. In daily interactions, people make privacy decisions often based on heuristics, shortcuts, feelings, and emotions. Such heuristics can prove quite successful most of the time — but can also lead to suboptimal behaviors or regrettable mistakes. Behavioral economics and decision research have analyzed which decision making hurdles individuals face when making privacy decisions and how emotions and cognition can influence disclosure behavior. More recently, a growing body of work has started examining how interfaces and technology could be designed to counteract biases responsible for regrettable privacy decisions.

- *Semi-automating privacy settings*. Machine learning techniques have been used to help derive privacy profiles that can be used to significantly reduce the number of privacy decisions individuals have to make and the number of privacy settings users have to configure manually. User-oriented machine learning techniques can also be developed to help users refine their privacy settings, leveraging user feedback to suggest modifications to these settings. Additional research in this area may result in approaches to privacy settings configuration that is significantly less burdensome and more accessible to users.

- *Semi-automating understanding of privacy policies.* Machine learning, natural language processing, and crowdsourcing are being used to develop techniques aimed at semi-automatically understanding website privacy policies. Such techniques offer the prospect of automatically (or semi-automatically) understanding key aspects of a natural language privacy policy and summarizing its most salient elements to users, possibly in a personalized fashion. The output of such functionality could also be used to populate formal models of privacy policies, which in turn could be used to verify compliance with relevant laws and regulations and identify inconsistencies with a site's actual practices.

Research in these areas requires interdisciplinary interaction between social scientists, human-computer interaction researchers, and machine learning researchers. It contributes most directly to our capabilities for understanding the social science of privacy and engineering of privacy.

G: Economics

Many of the benefits and harms of collecting, analyzing, sharing, and using data about individuals are economic in nature. For example, the Internet offers many tools for more efficient job matching (search engines, social networks, microblogging platforms), but these also increase the potential for labor market discrimination based on job-seekers' online presence. The lens of economics can help us in understanding such tensions and tradeoffs around privacy and in developing tools that can be employed to achieve socially desirable outcomes.

Concrete research questions around economics and privacy include the following:

- To what extent will advancements in data analytics and increasing amounts of consumer data be used to increase societal welfare, and to what extent, instead, will those technologies and data cause mainly a reallocation of economic surplus from data subjects to data holders?

- To what extent can privacy-enhancing technologies be deployed in manners that stimulate economic growth at the same time as allowing for privacy protection?

- What are the incentives of stakeholders (such as consumers, companies, and potential adversaries) that impact their privacy-relevant decisions and how can we affect those incentives through the design of economic mechanisms (markets), the introduction of privacy-enhancing technologies, and regulation?

- Can we augment the traditional tools of market design and mechanism design from game theory with notions and tools from privacy research, so that privacy concerns need not interfere with the behavior of agents or the social welfare attained by these economic systems?

Research in these areas can draw upon economics, behavioral science, computer science, and law and policy, and contributes most directly to our understanding of the social science of privacy and to policy for privacy.

H: Privacy and Society

Society's notion of privacy is related in laws, social practice, and the shape of the various institutions governing legal, social and cultural norms regarding personal data.  Therefore, our understanding of the nature of privacy, and its ongoing evolution, depends on research into the law, politics, philosophy, sociology and anthropology of privacy.  Research on the meaning and value of privacy should not merely consider choices relative to the status quo, but also be open to a complete reinvention of the means by which we store, share, buy, sell, track, compute on, and draw conclusions from potentially sensitive data.  At the same time, it should prepare us for a variety of future scenarios that may arrive through a combination of technological, social, political, and economic forces.  For example, if sensitive information about individuals becomes pervasively available, how can we ensure that it is used only in appropriate and fair ways?

A combination of disciplines is necessary in order to reflect the full range of society's privacy landscape.  Legal scholarship can contribute both to the historical foundations of privacy and to an understanding of how different regulatory mechanisms may be more or less effective for addressing certain privacy challenges.  Today's information privacy questions also require insight into the unique challenges technological change can pose to legal systems. Philosophy and social theory have already shown that they can offer valuable critical perspective on our cultural notions of privacy.  Political scientists can help understand the behavior of different governmental institutions with respect to privacy values.  And sociology, anthropology, together with those who specialize in Science, Technology and Society, all have vital roles to play in providing a well-grounded picture of the institutional, cultural role of privacy in human societies.

Research on privacy and society contributes most directly to our understanding of the social science of privacy and to policy for privacy.

**V. Fostering interdisciplinary work**

As the research directions in the previous section indicate, developing a science of privacy and effective privacy solutions requires a combined understanding of computing technology, information, human behavior, and governance mechanisms.  Thus, it is important to develop and strengthen structures that encourage interdisciplinary research on privacy.  We have a number of suggestions along these lines:

- There need to be more workshops and conferences where scholars and practitioners from different disciplines and domains can gather to exchange ideas, break down language barriers, and start collaborations.  The annual Privacy Law Scholars Conference (PLSC) is an excellent example, bringing law and policy researchers together with computer scientists and social scientists, but it is already stretched in its capacity and scope.  The Symposium On Usable Privacy and Security (SOUPS) fosters interaction between human-computer interaction researchers and security and privacy researchers. The Privacy Enhancing Technologies Symposium (PETS) solicits papers from many disciplines, but mostly receives submissions from researchers in the more technical privacy areas. The International Association of Privacy Professionals (IAPP) conferences attract mostly practitioners and few scholars. More venues are needed to foster interdisciplinary discussion.
- Designated research funding can incentivize and support interdisciplinary work. Workshops and seed funding can enable investigators to identify and develop potential collaborations in a bottom-up manner before having to submit a larger, long-term proposal.
- It is often difficult for researchers or practitioners to navigate the research literature in other disciplines. This can be remedied by encouraging more papers that systematize the results and unsolved problems from a particular area in a form that is accessible and understandable to outsiders, as well as identify problems that could benefit from other disciplines' insights. We also need more multidisciplinary publication venues that welcome such papers.
- The department-centered hiring and promotion process within academia can also present challenges for the career development of interdisciplinary researchers. Increased support for interdisciplinary research centers can help address this concern.

While interdisciplinary engagement is very important for privacy research, we should also recognize that many fundamental contributions to privacy originate within single disciplines and need time to mature before crossing disciplinary lines.  Thus, strengthened support for interdisciplinary research should not come at the expense of intradisciplinary work.

**VI. Transition to Practice**

Transition of privacy research to practice is critically important and likely to have a large positive impact on users, organizations, corporations, and governments.  This requires several types of efforts: engagement and education, bridging the gap between research results and working, deployed products, and creating funding models and structures that facilitate this transition.

Education and engagement should be directed at several different audiences:

- *System builders* who need to be integrating privacy technologies, algorithms, etc. into their apps, systems software, enterprise software, operating systems, data processing and data analysis systems, etc. To bring industry to the table, it is also important to explore organizational mechanisms that can help introduce privacy and data ethics into corporate decision making processes.  Coordination activities should help "bridge the chasm" between the research community and operational users of computing systems.  More systematization of research results can make them more accessible to potential users and help these users find appropriate technologies for their problems.
- *Regulators, policy makers, and lawmakers.*  In particular, it is important to make academic research outcomes accessible to policy makers and set appropriate expectations as to what is and is not achievable with the current state of the art in technology.  System designers also need help understanding whether designs satisfy privacy policy objectives and legal requirements.
- *Students (our future workforce).*  Workforce education is one of the key mechanisms for disseminating research results and influencing privacy practices in industry and government.  Basic privacy engineering skills should be taught to all undergraduate and graduate computer science students.  In addition, specialized graduate programs in privacy engineering can help prepare students for roles as privacy engineers.
- *The public.*  The public needs to be educated about privacy risks, as well as the capabilities and limitations of privacy protection technologies.  And, system designers should be supported in the application of HCI techniques that will help them understand what individual users and society at large needs from systems that handle personal data.

For bridging the gap between research results and deployed products, we need more funding models that encourage collaboration between privacy researchers and practitioners in specific application domains (such as those described in Section II).  Such translational research should be supported in addition to (rather than in place of) the many basic research efforts that are also needed to achieve our privacy objectives.

**Appendix**

A committee of five computer science researchers wrote this document in April of 2015. It is a research road map for the privacy community generated from previous reports on the topic of privacy and individual and group responses to a CCC request to the community that was made in February of 2015 and a Request for Information (RFI)-National Privacy Research Strategy that was made in September of 2014. The Computing Community Consortium edited the report. Special thanks to Tal Rabin and Salil Vadhan for leading this effort.

Report Writers
Lorrie Cranor, Carnegie Mellon University
Tal Rabin, IBM Research
Vitaly Shmatikov, Cornell Tech and The University of Texas at Austin
Salil Vadhan, Harvard University
Daniel Weitzner, Massachusetts Institute of Technology

Report Editors
Ann Drobnis, Computing Community Consortium
Helen Wright, Computing Community Consortium

Community Contributors
Alessandro Acquisti, Carnegie Mellon University
Yuvraj Agarwal, Carnegie Mellon University
Tawfig Alashoor, Georgia State University
Robert D. Atkinson, Information Technology and Innovation Foundation
Khaliah Barnes, EPIC Administrative Law Counsel
Richard Baskerville, Georgia State University
Lujo Bauer, Carnegie Mellon University
Elisa Bertino, Purdue University
Avrim Blum, Carnegie Mellon University
Travis Breaux, Carnegie Mellon University
Jan Camenisch, IBM Research
Daniel Castro, Information Technology and Innovation Foundation
Sarah A. Chauncey, Rockland BOCES and Learning Technologies Consultant
Scott Coull, RedJack
Anupam Datta, Carnegie Mellon University
Edward W. Felten, Princeton University
Steve Fienberg, Carnegie Mellon University
Christina Fong, Carnegie Mellon University
Seda Gurses, New York University
Andreas Haeberlen, University of Pennsylvania
Shai Halevi, Massachusetts Institute of Technology
Joe Hall, Center for Democracy and Technology
Moritz Hardt, IBM Research
Michael Hay, Colgate University

Julia Horwitz, Electronic Privacy Information Center
Jean-Pierre Hubaux, Ecole Polytechnique Fe'de'rale de Lausanne
Joanna N. Huey, Princeton University
J. Trevor Hughes, iapp
Farnam Jahanian, Carnegie Mellon University
Limin Jia, Carnegie Mellon University
Ari Juels, Cornell University
Murat Kantarcioglu, University of Texas at Dallas
Erin Kenneally, Elchemy
Jennifer King, University of California, Berkeley
Alfred Kobsa, University of California, Irvine
Alethea Lange, Center for Democracy and Technology
Insup Lee, University of Pennsylvania
Jae Ung Lee, SUNY Buffalo
Katrina Ligett, California Institute of Technology
Ashwin Machanavajjhala, Duke University
David P. Maher, Intertrust
Brad Malin, Vanderbilt University
Alan McQuinn, Information Technology and Innovation Foundation
H. Patricia McKenna, AmbientEase
Lee W. McKnight, Syracuse University
Gerome Miklau, University of Massachusetts
Deirdre K. Mulligan, University of California, Berkeley
Arvind Narayanan, Princeton University
Helen Nissenbaum, New York University
Dusko Pavlovic, University of Hawaii
Jon Peha, Carnegie Mellon University
Benjamin Pierce, University of Pennsylvania
Ann Racuya-Robbins, IT Developer World Knowledge Bank
H. Raghav Rao, SUNY Buffalo
Thomas Ristenpart, University of Wisconsin, Madison
Aaron Roth, University of Pennsylvania
Marc Rotenberg, The Electronic Privacy Information Center
Tuomas Sandholm, Carnegie Mellon University
Norman Sadeh, Carnegie Mellon University
Jeramie Scott, EPIC National Security Counsel
Doug Sicker, Carnegie Mellon University
Adam Smith, Pennsylvania State University
Oleg Sokolsky, University of Pennsylvania
Omer Tene, University of California, Berkeley
Bhavani Thuraisingham, University of Texas at Dallas
Jon Ullman, Columbia University
Rohit Valecha, SUNY Buffalo
John M. Willis, pINFOSEC
Rebecca Wright, Rutgers University

Ruilin Zhu, Georgia State University

*For citation use*: Cranor L., Rabin T., Shmatikov V., Vadhan S., & Weitzner D. (2015). *Towards a Privacy Research Roadmap for the Computing Community*: A white paper prepared for the Computing Community Consortium committee of the Computing Research Association.
http://cra.org/ccc/resources/ccc-led-whitepapers/

This material is based upon work supported by the National Science Foundation under Grant No. (1136993). Any opinions, findings, and conclusions or recommendations expressed in this material are those of the author(s) and do not necessarily reflect the views of the National Science Foundation.